\documentclass[prb,twocolumn,preprintnumbers,amsmath,amssymb,superscriptaddress,showpacs]{revtex4}
\usepackage{epsfig}
\usepackage{color}
\usepackage{graphicx}
\usepackage{dcolumn}
\usepackage{bm}

\newcommand{\be}{\begin{equation}}
\newcommand{\ee}{\end{equation}}
\newcommand{\bea}{\begin{eqnarray}}
\newcommand{\eea}{\end{eqnarray}}
\newcommand{\ket}[1]{\ensuremath{\left| #1 \right\rangle}}

\newcommand{\uj}{\ensuremath{^{j}}}
\newcommand{\uk}{\ensuremath{^{k}}}
\newcommand{\ul}{\ensuremath{^{l}}}
\newcommand{\defBy}{\equiv}

\newcommand{\science}{Science}
\newcommand{\nature}{Nature (London)}
\newcommand{\njp}{New J. Phys.}

\begin{document}

\title{Stability of longitudinal coupling for Josephson charge qubits}

\author{Carsten Hutter}
\affiliation{Institut f\"ur Theoretische Festk\"orperphysik
and DFG-Center for Functional Nanostructures (CFN),
Universit\"at Karlsruhe, D--76128 Karlsruhe, Germany}
\affiliation{Department of Physics, Stockholm University,
AlbaNova University Center, SE--106 91 Stockholm, Sweden}
\author{Yuriy Makhlin}
\affiliation{Landau Institute for Theoretical Physics,
Kosygin st. 2, 119334 Moscow, Russia}
\author{Alexander Shnirman}
\affiliation{Institut f\"ur Theoretische Festk\"orperphysik
and DFG-Center for Functional Nanostructures (CFN),
Universit\"at Karlsruhe, D--76128 Karlsruhe, Germany}
\author{Gerd Sch\"on}
\affiliation{Institut f\"ur Theoretische Festk\"orperphysik
and DFG-Center for Functional Nanostructures (CFN),
Universit\"at Karlsruhe, D--76128 Karlsruhe, Germany}

\begin{abstract}
For inductively coupled superconducting quantum bits, we determine
the conditions when the coupling commutes with the single-qubit
terms. We show that in certain parameter regimes such longitudinal
coupling can be stabilized with respect to variations of the
circuit parameters. In addition, we analyze its stability against
fluctuations of the control fields.
\end{abstract}


\maketitle

Among potential realizations of scalable quantum
computers~\cite{book2000nielsen}, superconducting qubits belong to
the promising
candidates\cite{rmp2001makhlin,bookchap2003devoret,bookchap2006wendin}.
Experimentally, coupling between quantum bits has been
demonstrated~\cite{nature2003pashkin,nature2003yamamoto,prl2004izmalkov,prl2005majer,science2005mcdermott},
as well as coupling of a qubit to an
oscillator~\cite{nature2004wallraff,nature2004chiorescu}. So far
most experiments have been performed with fixed coupling, but for
flexible and precise coherent control of a larger number of qubits,
a  tunable coupling is preferable. First demonstrations of such
tunable coupling have been provided only
recently~\cite{condmat2006vanderploeg,science2006hime}.

Although it is not necessary, for convenience and precision of the
operations, one wishes to be able to switch single-qubit and
coupling terms on and off independently. On the other hand, in
many physical realizations of qubits, coupling terms and
single-qubit terms are switched by the same control parameters.
Especially in designs which couple qubits via an
LC-oscillator\cite{nature1999makhlin,prl2002you}, the coupling
terms are present only when simultaneously single-qubit terms are
turned on. If this cannot be avoided, one should at least make
efforts to ensure that the coupling and single-qubit terms commute
with each other, a situation which we denote as ``longitudinal''
coupling. In this case, the time-evolution operator factorizes
into parts due to the coupling and due to single-qubit terms. If
the coupling can be switched off, one can easily undo unwanted
single-qubit operations and thus produce a ``pure'' two-qubit
gate. Moreover, if one can control the longitudinal-coupling
strength, the resulting operation is sensitive only to the time
integral and not to the detailed profile of the control pulse, a
property which makes operation easier and more stable.

Here, we investigate tunable longitudinal-coupling schemes for
Josephson charge qubits. At the charge degeneracy point, where
dephasing is weakest, capacitive (charge-charge) coupling is
always ``transverse'' (i.e. $\sigma_z \cdot \sigma_z$ coupling,
while the single qubit term is proportional to $\sigma_x$). Also
the tunable inductive (current-current) coupling scheme proposed
in Refs.~\onlinecite{nature1999makhlin,rmp2001makhlin}, where the
qubits are coupled to a common inductor, is purely transverse.
However, this scheme can be modified in such a way that the
coupling at the symmetry point becomes purely
longitudinal~\cite{prb2001you,prl2002you,prb2004lantz}.

The longitudinal-coupling designs proposed in
Refs.~\onlinecite{prb2001you,prl2002you,prb2004lantz} assume
identical junctions within each qubit. As we will show below,
variations of the junction parameters (especially critical
currents and capacitances), which are unavoidable due to
fabrication errors, add transverse terms to the coupling, which
can not be tuned to zero. This problem has also been noted in
Ref.~\onlinecite{njp2005wallquist}. Here, we propose and analyze a
design with longitudinal inductive coupling, which is stable
against such imperfections, in the sense that by tuning control
parameters, we can reach a point with purely longitudinal coupling.

The simplest method to compensate for variations in the critical
currents of the junctions in the circuit is to replace them by
dc-SQUIDs or more complicated circuits, which allow tuning of the
effective Josephson couplings via applied, individually controlled
magnetic fluxes. However, in order to reach effectively identical
junctions, one also needs tunable capacitances, for which there is
no simple and efficient solution. Nevertheless, we found a design
where, in spite of the spread in fabrication parameters, it is
possible to achieve longitudinal coupling merely by tuning fluxes
to specific values. For designs with two junctions per qubit
(those of Refs.~\onlinecite{prb2001you,prl2002you,prb2004lantz}
can be viewed as such), this is possible even when the Josephson
energies of these junctions are different, provided that, in
addition, the capacitances are different. In other words, the
system should deliberately be fabricated asymmetric.

Further, we analyze the stability with respect to time-dependent
fluctuations of the magnetic fluxes. Within the region of
parameters, where tuning to purely longitudinal coupling is
possible, the instability against fluctuations of fluxes increases
with increasing asymmetry of the capacitances. Thus, on one hand,
the system should be fabricated asymmetric in order to allow
tuning to longitudinal coupling; on the other hand, in order to
minimize the instability with respect to fluctuations in the
applied fluxes, one should keep the values of the capacitances as
close to each other as the first requirement permits.

\begin{figure}[bthp]
\centerline{
    \hspace{0.2cm}
    \epsfysize=36mm
    \epsfbox{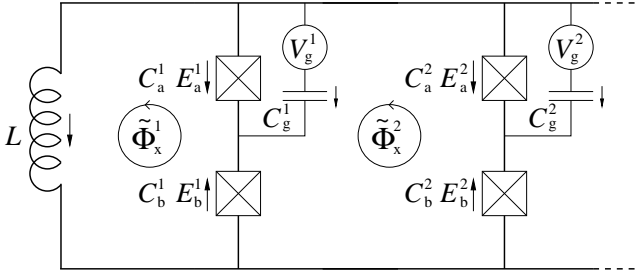}
  }
  \caption{\label{fig:outerFluxDesign}
  Charge qubits coupled to an $LC$-circuit. Earlier proposals with either longitudinal\cite{prl2002you} or
  transverse\cite{nature1999makhlin,rmp2001makhlin} coupling can be considered as particular cases of this design.
  The fluxes shown here are applied in the loops between the qubits, such that the total flux bias between the
  inductance and a qubit $j$ is given as $\Phi_{\rm x}\uj=\sum_{k=1}^j \tilde\Phi_{\rm x}^k$.
  The junctions may be replaced by SQUID loops to achieve tunable Josephson energies.
}
\end{figure}

{\it The system. }
As a specific example, we consider inductively coupled charge qubits as shown in Fig.~\ref{fig:outerFluxDesign}. For
particular values of the Josephson energies $E_{\rm a}\uj$ and capacitances $C_{\rm a}\uj$, it reduces to the design of
Refs.~\onlinecite{nature1999makhlin,rmp2001makhlin} with transverse coupling or to that of Ref.~\onlinecite{prl2002you}
with longitudinal coupling. The coupling can be controlled by fluxes applied in the loops. We use an upper index to
enumerate qubits and a lower index a/b to distinguish junctions above and  below the island of each qubit.

If $N$ qubits are coupled to the inductor $L$, there are $N+1$
independent phase variables in the system. As one of them, we
choose the phase drop $\phi_L=2\pi\Phi_L/\Phi_0$, related to the
flux $\Phi_L$ through the inductor. Furthermore, for each qubit
$j$, we introduce the phase $\phi\uj$ across the junction above the
qubit island. The system is then described by the Hamiltonian \bea
\label{eq:exactOriginal}
\hspace{-0.2cm}H&\hspace{-0.3cm}=\hspace{-0.4cm}& \sum_{j=1}^N
  \frac{\left(2en\uj-q\uj\right)^2}{2C_{\Sigma }\uj}
+\frac{1}{2C_s} [
    Q+\sum_{j=1}^N k_{\rm a}\uj (2en\uj-q\uj)
]^2
 \nonumber\\
&&\hspace{-0.7cm}+\frac{\Phi_L^2}{2L}-\sum_{j=1}^N
  E_{\rm a}\uj\cos\phi\uj
-\sum_{j=1}^N
 E_{\rm b}\uj\cos\left(\phi\uj-\phi_L-\phi_{\rm x}\uj\right)\ .
\eea Here $n\uj$ is the number of excess Cooper-pair charges on
the island of qubit $j$, canonically conjugate to $\phi\uj$. The
conjugate variable to $\Phi_L$ is the charge $Q$ on the capacitors
of the lower junctions of all qubits. The gate charges of the
qubit are defined as $q\uj=C_{\rm g}\uj V_{\rm g}\uj$, where
$V_{\rm g}\uj$ is the gate voltage and $C_{\rm g}\uj$ is the
corresponding capacitance. For simplicity, we consider constant
applied fluxes. The total qubit charge, screened by capacitances
with ratios $k_{\rm a}\uj\equiv C_{\rm b}\uj/C_\Sigma\uj$, acts as
gate charge of the oscillator. The total capacitance of each
island is $C_{\Sigma}\uj=C_{\rm a}\uj+C_{\rm b}\uj+C_{\rm g}\uj$.
The Josephson coupling terms of the Hamiltonian contain the total
phase bias between each qubit and the inductor, $\phi_{\rm
x}\uj=\sum_{k=1}^{j}\frac{2\pi}{\Phi_0}\tilde{\Phi}_{\rm x}\uj$.
We further introduced the total capacitance of the $LC$-circuit,
$C_{s}=\sum_{j=1}^N (C_{\rm a}\uj+C_{\rm g}\uj)C_{\rm
b}\uj/C_{\Sigma }\uj$.

We consider an oscillator with frequency much higher
than all qubit frequencies, $\hbar/\sqrt{LC_s}\gg E_{C}^j,\,E_{\rm
a}^j,E_{\rm b}^j$, and further assume $\sqrt{L/C_s}\ll \hbar/e^2$.
This ensures that both the average value and the fluctuations of
the oscillator flux $\Phi_L$ are much smaller than $\Phi_0$, and
the oscillator can be adiabatically eliminated, cf.
Ref.~\onlinecite{rmp2001makhlin}.
Below, we further discuss the condition of adiabaticity.
In the process
we transform the qubit phase as $\phi^j\rightarrow
\tilde{\phi}^j=\phi^j- k_{\rm a}^j \phi_L$. We further consider
the charge regime $E_{C}\uj =e^2/(2C_{\Sigma }\uj)\gg E_{\rm
a}\uj,E_{\rm b}\uj$. If for each qubit the gate voltage is chosen
close to a charge degeneracy point, we can employ a two-state
approximation with relevant charge states denoted as $\ket{n=1}$
and $\ket{n=0}$. After a transformation
$\tilde{\phi}\uj\rightarrow \theta\uj=\tilde{\phi}\uj-\phi_{\rm
x}\uj/2$ and the gauge choice $\cos\theta\uj\rightarrow
\sigma_{x}\uj/2$ and $\sin\theta\uj\rightarrow \sigma_{y}\uj/2$,
we find the effective Hamiltonian \be\label{eq:H_eff} H_{\rm
eff}=\sum_{j=1}^N H_{{\rm
single}}\uj-\frac{L}{2}\left(\sum_{j=1}^N I\uj\right)^2\ . \ee The
single-qubit Hamiltonian of qubit $j$ and its contribution to the
current through the inductor are given by \bea H_{{\rm
single}}\uj&=&-\frac{1}{2}\left(B_{x}\uj\sigma_{x}\uj+B_{y}\uj\sigma_{y}\uj
+B_{z}\uj\sigma_{z}\uj\right)\,,\nonumber\\
I\uj&=&\frac{1}{2}\left(I_{x}\uj\sigma_{x}\uj+I_{y}\uj\sigma_{y}\uj\right)\,,
\eea
where
\bea
\label{eq:Bs}
B_{x}&=&(E_{\rm a}+E_{\rm b})\;\cos(\phi_{\rm x}/2)\,,
\nonumber\\
B_{y}&=&(-E_{\rm a}+E_{\rm b})\;\sin(\phi_{\rm x}/2)\,,
\nonumber\\
B_{z}&=&4E_{C}(q/e-1)\,.
\eea
In Eqs.~(\ref{eq:Bs})
(and below, where it does not lead to confusion) we omit
the qubit index $j$. The current of each
qubit is
\bea
\label{eq:Is}
I_{x}&=&[k_{\rm a}\cdot I_{\rm a}+(1-k_{\rm a}) \cdot I_{\rm b}]\;\sin(\phi_{\rm x}/2)
\nonumber\\
I_{y}&=&[k_{\rm a}\cdot I_{\rm a}+(k_{\rm a}-1) \cdot I_{\rm
b}]\;\cos(\phi_{\rm x}/2)\ , \eea where the critical currents of
the junctions above and below the island of a qubit are denoted by
$I_{\rm a/b}=2\pi E_{\rm a/b}/\Phi_0$.

The design of Ref.~\onlinecite{prl2002you} with longitudinal coupling is recovered for symmetric Josephson energies
and capacitances ($E_{\rm a}=E_{\rm b}$ and $k_{\rm a}=1/2$). On the other hand, the coupling is always
transverse\cite{prl1997shnirman} for $E_{\rm a}=0$ or $E_{\rm b}=0$.

{\it Conditions for longitudinal coupling.} We will determine now
under which conditions the coupling is longitudinal, that is, the
single-qubit terms commute with the coupling term, \be
    [\ H_{\rm single}\uj\,\, ,
      \,\,L \sum\nolimits_{k,l}I\uk I\ul \ ] = 0 \quad\mathrm{for\
      all\ } j\, .
\label{eq:commzero} \ee First, we assume operation at the charge
degeneracy point, $B_{z}\uj=0$, of each qubit $j$. Here
decoherence is slowest, and only the tunneling energy part,
$U\uj=-(1/2)B_{x}\uj\sigma_{x}\uj-(1/2) B_{y}\uj\sigma_{y}\uj$, of
the single-qubit energy remains. Since the current operators of
different qubits commute, the condition (\ref{eq:commzero}) is
satisfied if and only if $[I\uj, U\uj]=0$ for each qubit $j$, and
it is sufficient to consider each qubit separately. This
commutator vanishes when the following condition is fulfilled for
each qubit \bea \label{eq:newcondition} k_{\rm a} E_{\rm
a}^2+(k_{\rm a}-1)E_{\rm b}^2 +(2k_{\rm a}-1)E_{\rm a} E_{\rm
b}\cos\phi_{\rm x}=0\ . \eea As expected, this condition of
longitudinal coupling is $2\pi$-periodic in the applied phases.
Apart from trivial cases, where either single-qubit and coupling
terms vanish ($E_{\rm a}=E_{\rm b}=0$), or where the connection on
one side of the island is broken ($E_{\rm a}=0$ and $k_{\rm a}=1$
or $E_{\rm b}=0$ and $k_{\rm a}=0$), this condition can only be
fulfilled in the following two cases: (i) $E_{\rm a}=E_{\rm b}$
and $k_{\rm a}=1/2$, or (ii)
\begin{equation}
  \label{eq:condition}
  \cos\phi_{\rm x}=\frac{\kappa-\epsilon^2}{(1-\kappa)\epsilon}\ .
\end{equation}
Here we introduced the ratios of Josephson couplings and of
capacitances above and below the junction, \be
\label{eq:defEpsKappa} \epsilon\defBy\frac{E_{\rm a}}{E_{\rm
b}}\quad,\quad \kappa\defBy\frac{1-k_{\rm a}}{k_{\rm a}}
=\frac{C_{\rm a}+C_{\rm g}}{C_{\rm b}}\ . \ee

Note that case (i) corresponds to the symmetric situation: equal
Josephson couplings and capacitances on both sides of the island
($\epsilon=\kappa=1$). While the former may be adjusted by
replacing junctions by SQUID loops, the spread in the values of
the capacitances can not be avoided in real experiments.

\begin{figure}[bthp]
  \centerline{
\hspace{0.2cm}
    \epsfysize=50mm
    \epsfbox{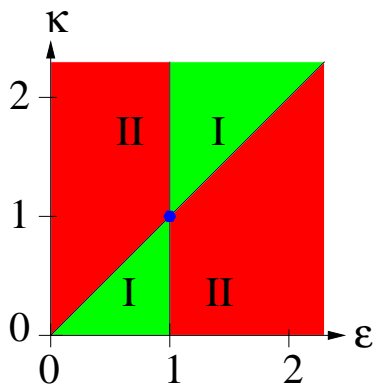}
  }
  \caption{  \label{fig:phaseDiagram}
  (Color online) Stability diagram for a Josephson charge qubit. In region I tuning to longitudinal coupling by
  applied magnetic fluxes is possible. In region II the coupling always has a transverse component.
}
\end{figure}

{\it Tuning to longitudinal coupling.} We will now show that for
case (ii) for a wide range of design parameters -- in spite of
some spread -- one can tune to longitudinal coupling by proper
choice of the applied flux for each qubit. Especially,
condition~(\ref{eq:condition}) can be satisfied even when
capacitances and Josephson energies of the fabricated setup
deviate from their nominal values. Via the applied fluxes in the
design of Fig.~\ref{fig:outerFluxDesign}, the phases $\phi_{\rm x}$
can be tuned in the full range between $0$ and $2\pi$, and the
condition~(\ref{eq:condition}) can be satisfied, when \be
    \label{eq:parameterRegime}
    1)\ \kappa \leq \epsilon \leq 1
    \quad \mbox{or} \quad
    2)\ 1 \leq \epsilon \leq \kappa \,.
\ee This parameter range, where it is possible to tune to
longitudinal coupling, is shown in Fig.~\ref{fig:phaseDiagram} as
region~I (green). On the other hand, in region~II of this diagram,
purely longitudinal coupling can not be achieved.

Only for the symmetric case, $\kappa=\epsilon=1$, which
corresponds to a single point in the diagram shown in
Fig.~\ref{fig:phaseDiagram}, the coupling is longitudinal for an
arbitrary value of the parameter $\phi_{\rm x}$. For this reason,
the symmetric design appears most appealing. However, it is
unstable in the sense that small deviations of the parameters
$\epsilon$ or $\kappa$ from the nominal value of $1$ may bring the
system into region II, where longitudinal coupling cannot be
achieved. Therefore, in order to be able to tune to stable
longitudinal coupling, one should aim for design parameters inside
region I of the phase diagram. Since the commutation relation
should be satisfied for all qubits, one needs to build each qubit
with parameters in the proper range. Further, one needs to have
for each qubit an individual control parameter $\phi_{\rm x}$ (as
opposed to a common flux through the inductor used in
Ref.~\onlinecite{prl2002you}).

{\it Stability of longitudinal coupling with respect to
fluctuations in the control fluxes.} Time-dependent fluctuations
or drifts in the applied fluxes will shift the operation point
from the desired values given by Eq.~(\ref{eq:condition}). We want
to estimate for which values of the design parameters these
deviations lead to the weakest instability of the longitudinal
coupling. For this purpose we introduce a measure for the residual
transverse-coupling strength. If one regards the single qubit
terms at the degeneracy point and the current of each qubit as
vectors, $(B_x,B_y)$ and $(I_x,I_y)$, in the $x$-$y$ plane,
the commutation between both corresponds to parallel vectors. On the
other hand, the $z$-component of their cross product, $B_x I_y-B_y
I_x$, which is also proportional to the lhs. of
Eq.~(\ref{eq:newcondition}), is a measure for the contribution of
one qubit to the transverse coupling. To gain independence of the
total coupling strength, we consider the ratio of transverse and
longitudinal coupling strengths, \bea t=\frac{B_x I_y-B_y I_x}{B_x
I_x + B_y I_y}\ . \eea The numerator and denominator vanish for
purely longitudinal or transverse coupling, respectively. Larger
values of $|t|$ correspond to stronger transverse coupling. We
find \bea \hspace{-0.5cm} t&=&\frac{k_{\rm a} E_{\rm a}^2+(k_{\rm
a}-1)E_{\rm b}^2+(2k_{\rm a}-1)E_{\rm a}E_{\rm b}\cos\phi_{\rm x}
    }{E_{\rm a}E_{\rm b}\sin\phi_{\rm x}}\ .
\eea When $\phi_{\rm x}=\phi_{{\rm x}0}+\delta\phi_{\rm x}$
deviates from the value $\phi_{{\rm x}0}$ satisfying the
condition~(\ref{eq:condition}), the variation of the relative
coupling strength in the linear order is governed by \bea
\left.\frac{\partial t}{\partial (\delta\phi_{\rm
x})}\right|_{\delta\phi_{\rm x}=0}=1-2k_{\rm a}
=\frac{\kappa-1}{\kappa+1}\,. \label{eq:dtdf} \eea Note that this
expression holds in region~I of the stability diagram for the
special values of the flux bias~(\ref{eq:condition}), at which $t$
vanishes. Thus, on one hand, to tune to longitudinal coupling one
has to aim inside region~I, not too close to its boundary; on the
other hand, in order to minimize the sensitivity to
fluctuations~(\ref{eq:dtdf}), one should keep the value of
$\kappa$ close to $1$.

{\it Slow and fast oscillators.} So far we
considered a fast oscillator, $\hbar\omega_{LC}=
\hbar/\sqrt{LC_s}\gg E_{\rm qb} \sim E_{C}^j,\,E_{\rm a}^j,E_{\rm b}^j$, which
could be adiabatically eliminated. Let us briefly discuss the opposite case of a slow
oscillator (cf.~Ref.~\onlinecite{pra2000soerensen}; note that $\omega_{LC}$
decreases with the number $N$ of qubits).
For this purpose, it is convenient to split the operators $I^j$ into
longitudinal and transverse parts, $I^j = I^j_{\rm long}+I^j_{\rm
trans}$. Since $I^j_{\rm long}$ is a slow variable (as long as the single-qubit
Hamiltonian is kept fixed), the longitudinal coupling
term is of the same form as above, $-{\textstyle\frac{L}{2}}\left(\sum_{j=1}^N
I_{\rm long}\uj\right)^2$. Notably, the transverse coupling is suppressed:
$I_{\rm trans}$ varies at
frequencies $\sim E_{\rm qb}\gg \omega_{LC}$, at which the response of the
oscillator is weak, and this results in a transverse inductive coupling
suppressed by a factor $(\omega_{LC}/E_{\rm qb})^2$ relative to the longitudinal coupling.
Thus, in case of a slow oscillator the longitudinal coupling is even more stable.
This description of the coupling
applies on time scales longer than $\omega_{LC}^{-1}$. This constraint does not
allow a decrease in $\omega_{LC}$ indefinitely and, in particular, limits the
scalability, i.e., the number $N$ of qubits.

{\it Comments.} (i) Besides working in the parameter regime given
in Eq.~(\ref{eq:parameterRegime}), one should keep the gate
capacitances $C_{\rm g}$ small, in order to decouple the qubits
from the electromagnetic environment. These can be conflicting
requirements, since the ratio of Josephson energies $E_{\rm
a}/E_{\rm b}$ is usually close to that of the capacitances $C_{\rm
a}/C_{\rm b}$ in a fabricated sample. This problem can be lifted
by employing an additional capacitance in parallel to either of
the qubit junctions.

(ii) We assumed individually controllable fluxes in qubit loops.
In experiment, fluxes are controlled via multiple current lines.
However, cross couplings may appear, which can be
overcome\cite{science2005mcdermott}.

(iii) For the design as discussed so far, there is only one
control flux for each qubit which is used to tune to the point of
longitudinal coupling. Since one would like to control the
coupling strength independently, additional control parameters are
desirable. One possibility is to replace simple junctions by SQUID
loops with additional control fluxes. These can be used to change
the effective Josephson energies and therefore the coupling
strength. Furthermore, when one keeps the ratio of effective
Josephson energies constant, one can stay at the point of
longitudinal coupling, while changing the longitudinal coupling
strength. We emphasize the difference to the proposal of
Ref.~\onlinecite{prl2002you}, where only one common flux in
``outer loops" is applied, and longitudinal coupling can not be
achieved in this way for asymmetric design parameters.

(iv) Alternatively, tunable coupling strength may be achieved by a
standard replacement of the inductor with a Josephson junction in
the phase regime, see, e.g., Ref.~\onlinecite{rmp2001makhlin}. The
tunability can be achieved by current-biasing this
junction~\cite{prb2004lantz}. This has an additional advantage
since in an experiment currents can be switched faster than
fluxes. In order to produce longitudinal coupling in such a
situation, the condition of Eq.~(\ref{eq:condition}) needs to be
slightly modified; one has to add the additional phase applied by
the control current to $\phi_{\rm x}$ for each qubit.

(v) Searching for longitudinal coupling which is stable against
weak flux fluctuations, one may consider more involved designs, in
particular, with Josephson junctions replaced by dc-SQUID loops or
even more junctions in parallel. This would provide more tuning
parameters (magnetic fluxes in the loops) and potentially the
possibility to find a better operation point. However, such a
point should be stable against fluctuations in all control fluxes.
Moreover, the additional fluxes will also fluctuate and increase
the noise level. One can show that it is impossible to stabilize
the longitudinal coupling in this manner~\cite{thesis}.

{\it In summary,} we analyzed, for a system of inductively coupled
charge qubits, the conditions for longitudinal inter-qubit
coupling, i.e., a coupling that commutes with single-qubit terms.
Earlier suggestions relied on precise symmetries of parameters of
various junctions. We studied the stability of the longitudinal
coupling with respect to deviations from this nominal symmetry,
which is unavoidable during fabrication. We have shown that in a
simple design one can reach longitudinal coupling by keeping the
nominal parameters in a certain range, away from the symmetry
point. For such a circuit, one can always compensate for
deviations of the parameters of the fabricated circuit by tuning
control fluxes. Fluctuations of magnetic fluxes render the
longitudinal character unstable for any circuit of the considered
type. We found the conditions when this instability is weak.

This work is part of the EU IST project EuroSQIP, and was also
supported by the RSSF (YM) and by Graduiertenkolleg "Kollektive
Ph{\"a}nomene im Festk{\"o}rper" (CH). We acknowledge valuable
discussions with M. Wallquist.

\end{document}